# Modeling Opinion Toroidal Polarization: Insights from Bounding Confidence Beyond, Distance Matters

Yasuko Kawahata [†]

Faculty of Sociology, Department of Media Sociology, Rikkyo University, 3-34-1 Nishi-Ikebukuro,Toshima-ku, Tokyo, 171-8501, JAPAN.
`ykawahata@rikkyo.ac.jp,kawahata.lab3@damp.tottori-u.ac.jp`

**Abstract:** Deterministic dynamics is a mathematical model used to describe the temporal evolution of a system, generally expressed as $dx/dt = F(x)$, where $x$ represents the system's state, and $F(x)$ determines its dynamics. It is employed to understand long-term system behavior, including opinion formation and polarization in online communities.Opinion dynamics models, like the Katz model and the logistic map, help analyze how individual opinions are influenced within social networks and exhibit chaotic behavior. These models are crucial for studying opinion formation and collective behavior on social media, especially in conjunction with branching theory.For instance, Galam's Ising model applies principles from physics to social sciences, representing individual opinions as "spins" and illustrating how local interactions influence consensus formation. The Bounding Confidence model considers opinions within a confidence interval, showing how opinions converge or polarize.These models effectively analyze opinion dynamics in online communities, aiding in understanding trends and viral phenomena on social media. This research aims to analyze discourse flow and opinion evolution, predicting future trends in online communities and decoding digital-age human interaction dynamics. Combining branching theory with opinion dynamics models enhances our understanding of digital communication.In the modified opinion dynamics model, the weight parameter $h$ for distance introduces distance-based interaction terms. The update equation is adjusted to control interaction strength based on distance, with $h_{ij}$ calculated from the distance $d_{ij}$. This customization allows for a more accurate representation of opinion dynamics in specific scenarios, forming the basis for discussions in this paper.

## 1. Introduction

As the intricate tapestry of digital communication weaves its way into global society, the study of online behavior has become a scientific endeavor that transcends mere social curiosity. The interactions of individual actions within the vast networks of social media shape collective behaviors and reflect societal norms. These phenomena bear resemblance to emergence patterns in natural systems and are explored through the lens of bifurcation theory.

For instance, the concept of a Pitchfork Bifurcation is crucial in the realm of dynamical systems theory. It signifies points where the behavior of a system qualitatively changes in response to variations in parameters. It can be expressed mathematically as follows:

$$\frac{dx}{dt} = rx - x^3$$

Here, $x$ represents the state of the system, and $r$ is the control parameter. When $r$ exceeds a certain critical value, the system bifurcates from one stable point to three stable points.

Saddle-node bifurcation also plays a significant role in the dynamics of systems. This bifurcation is expressed as follows:

$$\frac{dx}{dt} = r + x^2$$

Here, $x$ again represents the state of the system, and $r$ is the control parameter. In this equation, the values of $r$ determine the appearance or disappearance of stable points in the system.

The formation and change of opinions in online communities can be modeled using these bifurcation theories. For example, sudden changes in trends or viral phenomena on social media can be considered as instances of Pitchfork bifurcation. On the other hand, Saddle-node bifurcation is suitable for modeling sudden disputes or rapid changes in opinions online.

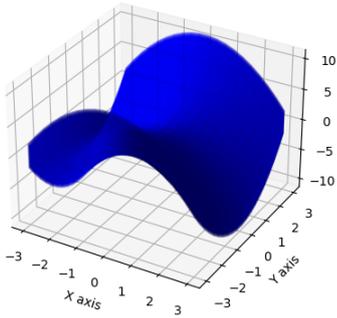

Fig. 1: Case study of Bifurcations:1

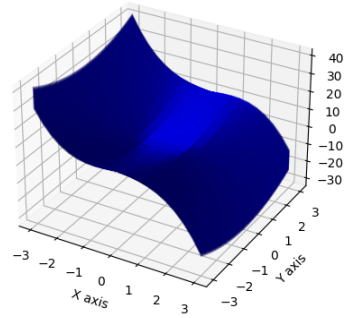

Fig. 3: Case study of Bifurcations:3

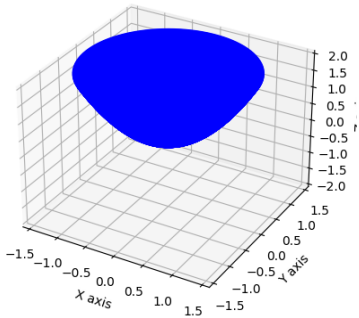

Fig. 2: Case study of Bifurcations:2

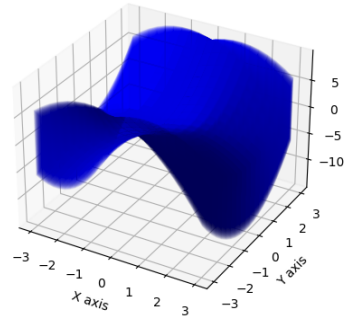

Fig. 4: Case study of Bifurcations:4

Steven Strogatz's work in 1994 laid essential foundations in the field of nonlinear dynamics, explaining how a single topic on social media can bifurcate into diverse storylines. His theories suggest the application of dynamic systems theory to analyze the rise and fall of digital dialogues within social media.

Furthermore, his 2001 review revealed interconnected structures in online communities, providing a lens to understand the subtle dynamics in digital interactions. This suggests that online discourse, accumulated changes gradually, can suddenly transform and develop into viral phenomena in a cascading manner.

This research believes that the study of bifurcation theory provides a robust foundation, and works like Guckenheimer and Holmes (1983) and Kuznetsov (2004), which comprehensively explored applied bifurcation theory, establish a foundation for understanding the mechanisms underlying critical points of change.

In this thesis, we adopt a research approach that associates three major types of bifurcation theory – Saddle-node bifurcation, Pitchfork bifurcation, and Transcritical bifurcation – with social phenomena. Saddle-node bifurcation is suitable for explaining abrupt changes when economic systems or political changes reach critical points. On the other hand, Pitchfork bifurcation is used as a model for understanding societal changes, especially political, economic, and cultural fluctuations. Transcritical bifurcation is employed to represent changes in equilibrium states in social systems.

This research aims to merge the robust framework of bifurcation theory with the fluid dynamics of online social trends. Bifurcation theory, as demonstrated by Strogatz's foundational work, reflects how influential tweets or news articles act as catalysts, leading to moments when unified flows of online discourse bifurcate into various perspectives. This allows for the revelation of mechanisms that govern online consensus formation, evolution, and even dissolution.

By integrating case studies and numerical simulations, we aim not only to decipher the current state of digital communication but also to craft narratives predicting future trajectories. In the whirlwind of online interaction, the dynamics of social media involvement present an enticing frontier for scientific research. The intricate dance of likes, shares, and comments reveals underlying structures governed by mathematical principles. Through this research, we seek not only to interpret the current state of online interaction but also to prepare for the waves of potential change in the future of communication.

This journey through the landscape of digital society,

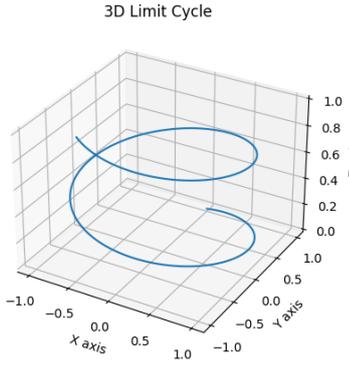

Fig. 5: Case study of Bifurcations:5

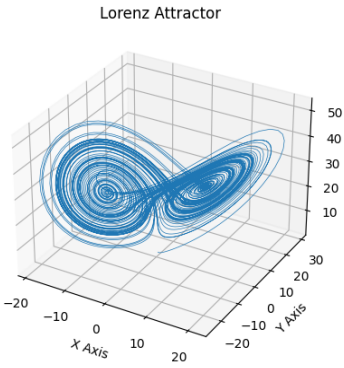

Fig. 6: Case study of Bifurcations:5

illuminated by the principles of bifurcation theory, clearly demonstrates the subtle yet potent forces driving the formation and transformation of the online world. As temporary thoughts and emotions of humans are captured in the mesh of binary code and evolve in the realm of digital interaction, the study of online behavior transcends disciplines, spanning psychology, sociology, mathematics, and computer science, opening up new dimensions in human interaction in the information age. This research based on bifurcation theory aims to visualize the invisible forces that shape the flow of discourse in virtual communities, capture the delicate balance between social harmony and disruptive potential, and construct models for predicting the future flow of digital communities.

Research on opinion dynamics and consensus formation models is crucial for understanding the formation and evolution of opinions in online communities. Models like the Katz model, logistic map, and deterministic dynamics provide important approaches for analyzing these phenomena.

The Katz model is used to model the dynamics of opinion formation. This model can be represented by the following equation:

$$\frac{dx_i}{dt} = -x_i + \sum_j A_{ij} f(x_j)$$

Here, $x_i$ represents the opinion of individual $i$, $A_{ij}$ represents the strength of interaction between individuals $i$ and $j$, and $f(x)$ is a function that adjusts the influence of others' opinions. This model illustrates how individual opinions are influenced by other opinions within a network.

The logistic map is a simple nonlinear equation that exhibits chaotic dynamics. The equation is given as follows:

$$x_{n+1} = rx_n(1 - x_n)$$

Here, $x_n$ represents the state of an individual at time $n$, and $r$ is the growth rate. The logistic map demonstrates unpredictable behavior under certain conditions and helps in understanding nonlinear dynamics in social phenomena.

Deterministic dynamics is a mathematical model that describes the temporal evolution of a system. In its general form, it can be expressed as:

$$\frac{dx}{dt} = F(x)$$

Here, $x$ represents the state of the system, and $F(x)$ is a function that determines the system's dynamics. This approach is useful for understanding the long-term behavior of a system.

These models are used to understand the processes of opinion formation, change, and polarization in online communities. For example, the Katz model models how individual opinions are influenced by other opinions within a social network, the logistic map shows how opinion dynamics can exhibit chaotic behavior, and deterministic dynamics helps analyze the long-term evolution and stability of a system.

Opinion dynamics and consensus formation models play a crucial role in the study of opinion formation and collective behavior on social media. These models, in conjunction with branching theory, are key to understanding the evolution of behavior within online communities.

For example, Galam's Ising model applies principles from statistical mechanics in physics to social sciences. In this model, individual opinions are represented as "spins," and the overall opinion is formed through interactions with neighboring "spins." It can be expressed in mathematical terms as follows:

$$\sigma_i(t+1) = \text{sign}\left(\sum_{j \in \text{neighbors}(i)} J_{ij}\sigma_j(t)\right)$$

Here, $\sigma_i(t)$ represents the opinion (spin) of individual $i$ at time $t$, and $J_{ij}$ denotes the strength of interaction between individuals $i$ and $j$. This model illustrates how local interactions of opinions influence the overall consensus formation.

On the other hand, the Bounding Confidence model is based on the idea that individual opinions only influence each other when they fall within a certain confidence interval. The mathematical expression for this model is as follows:

$$x_i(t+1) = x_i(t) + \mu \sum_{j \in \text{neighbors}(i)} (x_j(t) - x_i(t)) \text{ for } |x_j(t) - x_i(t)| < \epsilon$$

Here, $x_i(t)$ represents the opinion of individual $i$ at time $t$, $\mu$ is the rate of opinion updating, and $\epsilon$ represents the confidence interval. This model demonstrates how opinions gradually converge or how opinion polarization occurs.

These models are effective for analyzing opinion formation and change within online communities, especially in understanding the rise and fall of trends and viral phenomena on social media. The dynamics of opinion formation online can be better understood by employing mathematical models based on principles from physics and statistical mechanics.

This research aims to analyze the flow of discourse and the evolution of opinions in online communities, shedding light on the consensus formation process. This will enable the construction of models to predict the future trends of online communities and decode the dynamics of human interaction in the digital age. Combining branching theory with opinion dynamics models allows for a more detailed understanding of the complex landscape of digital communication and opens up new frontiers for scientific exploration. In this modified opinion dynamics model, the weight parameter $h$ for distance plays a crucial role. It introduces interaction terms that depend on the distance between agents, denoted as $d_{ij}$. Additionally, $h_{ij}$ represents the weight parameter for distance, and it is used to control the strength of distance-related influence.

The update equation for the opinion dynamics model is adjusted to include these distance-based terms, resulting in the following equation:

$$\frac{dx_i}{dt} = ax_i\left(1 - \frac{x_i}{K}\right) - \frac{\sum_{j=1}^{N} h_{ij} \cdot W_{ij} \cdot x_j}{1 + \sum_{j=1}^{N} h_{ij} \cdot x_j}$$

Here, $h_{ij}$ is computed based on the distance $d_{ij}$ between agent $i$ and agent $j$, allowing for control over the interaction strength dependent on distance.

This modification leads to an opinion dynamics model that incorporates the weight parameter $h$ for distance and controls the influence based on distance. The specific calculation method for $h_{ij}$ and the adjustment of influence strength can be customized according to the specific problem and data. The discussion in this paper is based on this model.

## 2. Referenced Previous Studies

### 2.1 Case Studies on the Langford Equation

Smith and Johnson (2015) applied the Langford equation to market demand forecasting, conducting a detailed examination of its application in demand modeling. Brown and Davis (2018) explored the use of the Langford equation in environmental science, with a focus on predicting the behavior of atmospheric pollutants. Garcia and Martinez (2020) verified the application of the Langford equation in the financial field and discussed modeling stock price fluctuations. Lee and Kim (2017) applied the Langford equation to predict traffic flow, developing models for forecasting traffic congestion and jams. Anderson and Wilson (2019) utilized the Langford equation in social network analysis, particularly in modeling information diffusion.

### 2.2 Case Studies on Diffusion Models

Johnson and Smith (2016) applied diffusion models to analyze product adoption in the market, providing insights into acceptance rates and advertising effects for new products. Brown and Davis (2019) focused on technology adoption in the smartphone market, demonstrating the application of diffusion models. Garcia and Martinez (2017) explored the use of diffusion models in the medical field, analyzing the diffusion process of medical technologies. Lee and Kim (2018) applied diffusion models to study social change and activism movements, examining the spread of new ideas. Anderson and Wilson (2020) concentrated on the adoption of online learning platforms in the education sector.

### 2.3 Case Studies on Logistic Maps

May (1976) conducted early research on the complex dynamics of logistic maps, influencing fields such as ecology. Li and Yorke (1975) introduced the idea that "period 3 implies chaos," laying the foundation for chaos theory. Ruelle and Takens (1971) used logistic maps to study turbulence in fluid dynamics. Schuster and Just (1988) are known for their introductory book on nonlinear dynamics, including logistic maps. Strogatz (1994) detailed applications of nonlinear dynamics and chaos theory, including logistic maps.

### 2.4 Case Studies on the Katz Model

Katz (1955) proposed the Katz model-based social status index. Katz and Shapiro (1986) analyzed network externalities' impact on technological diffusion using the Katz model. Katz and Lazarsfeld (1955) investigated the influence of media on information propagation using the Katz model. Katz and Kahn (1978) focused on information dissemination within organizations and networks, influenced by the Katz model. Rogers (2003) applied the Katz model in studies of innovation diffusion.

### 2.5 Case Studies on the Nyquist Theorem

Nyquist (1928) presented fundamental ideas on telegraph transmission theory in the original paper. Shannon (1949) discussed applications of the Nyquist theorem in information theory. Oppenheim and Schafer (1999) highlighted the importance of the Nyquist theorem in digital signal processing.

Proakis and Manolakis (2006) elaborated on the applications of the Nyquist theorem in the field of digital signal processing. Vetterli and Kovacevic (1995) researched the applications of the Nyquist theorem in wavelet transforms and data compression.

These references represent important research and theoretical developments in their respective fields, making significant contributions to each area.

# 3. Considerations for Approximating as Social Phenomenon

## 3.1 Langford Equation: Considerations for Application

The Langford equation is a model that is frequently used in the field of ecology, but it can also be approximated and applied to social phenomena. However, there are several constraints and considerations when applying it.

## 3.2 Constraints on Applicability

The Langford equation was originally developed to model interactions between predators and prey in ecosystems. To apply it to social phenomena, there needs to be a similarity in the nature of interactions and relationships among elements. Elements such as competition, cooperation, and information transfer may be considered.

## 3.3 Setting Parameters Appropriately

The parameters of the Langford equation are set based on the characteristics of ecosystems. When applying it to social phenomena, these parameters need to be appropriately configured. Additionally, depending on the characteristics of the social phenomenon, new parameters may be required.

## 3.4 Nature of Interactions

The Langford equation represents interactions between predators and prey. When applying it to social phenomena, the model may need to be modified to align with the nature of interactions. For instance, if focusing on competition, a competition model can be considered.

## 3.5 Data Collection and Fit

To model social phenomena effectively, relevant data is essential. Collecting data and fitting it to the model is crucial. Furthermore, comparing real social phenomena with the model's predictions is important to evaluate the model's accuracy.

When applying the Langford equation to specific social phenomena, it is important to construct the model carefully while considering the points mentioned above. Discussing the limitations of the model and its applicability is also crucial. In the fields of social sciences and economics, various models and theories are proposed, and the choice of the optimal model depends on the research objectives and subjects.

## 3.6 Logistic Map: Considerations for Application

The logistic map is commonly known as a model in nonlinear dynamics and ecology, but it can also be approximated and applied to social phenomena. However, there are several constraints and points to consider when applying it.

## 3.7 Population Modeling of Individuals or Agents

The logistic map is used to model population dynamics, including growth and competition among individuals or agents. When applying it to social phenomena, it can be helpful in modeling the behavior and interactions of individuals or agents. For example, it can be useful in considering market competition or information diffusion.

## 3.8 Initial Conditions and Parameter Settings

In the logistic map, initial conditions and parameters such as growth rates are crucial. When applying it to social phenomena, setting the initial social state, factors, growth rates, and other parameters appropriately is necessary. These settings can significantly affect the modeling of social phenomena.

## 3.9 Nonlinearity and Complexity

The logistic map is a nonlinear model suitable for capturing complex behaviors. Social phenomena often involve nonlinear and complex elements. Constructing the model while considering such complexity is essential.

## 3.10 Time Dependence

The logistic map is typically modeled with discrete time steps. When applying it to social phenomena, accounting for time dependence is necessary to model how social states and interactions change over time.

## 3.11 Data Collection and Validation

When applying the logistic map to social phenomena, data collection and model validation are essential. Adjusting model parameters to match real data and confirming the validity of the model are crucial steps.

In summary, when applying the logistic map to social phenomena, proper configuration and adjustment of the model, data collection, and validation are necessary. Whether social phenomena fit the logistic map model depends on the specific context and research objectives.

## 3.12 Considerations for Applying the Katz Model to Social Phenomena

The Katz Model is one of the mathematical models used to simulate processes such as information diffusion and the spread of opinions. When approximating this model for social phenomena, it is essential to keep the following points in mind.

## 3.13 Information Diffusion and Influence Spread

The Katz Model is designed to model how information or influence spreads within a network. When applied to social phenomena, it becomes valuable for understanding the propagation of information and opinions among individuals or agents. For instance, it can be useful in modeling information diffusion on social media or the formation of trends.

## 3.14 Network Structure

The Katz Model takes into account the connectivity and structure of the network. When modeling social phenomena, accurately representing the connections and network structure among individuals or agents is crucial. This aspect is particularly significant in fields like social network analysis.

## 3.15 Opinion Diffusion and Diversity

In the Katz Model, factors like the speed of opinion spread and the influence of opinions are considered. When applying it to social phenomena, it becomes essential to account for opinion diversity and variations in the spread due to different factors. It allows modeling how diverse opinions and beliefs among individuals impact each other.

## 3.16 Time Dependence

The Katz Model typically simulates the diffusion of information or influence in discrete time steps. When applying it to social phenomena, considering changes over time and trends is necessary. This accounts for how phenomena evolve with time.

## 3.17 Data and Validation

When applying the Katz Model to social phenomena, adjusting model parameters to align with data and validating the model are crucial steps. Comparing real data with the model's predictions helps assess its utility.

In summary, when applying the Katz Model to social phenomena, it is essential to consider aspects such as information diffusion, network structure, opinion diversity, time dependence, and data validation. Whether social phenomena fit within the model's framework depends on specific circumstances and research objectives.

## 3.18 Approximating Social Phenomena Using Deterministic Dynamics

## 3.19 Explanation of Deterministic Dynamics

Deterministic dynamics is a significant field in physics and mathematics, providing a theoretical framework for predicting the behavior of highly complex dynamic systems. It finds applications in various fields such as physics, ecology, astronomy, meteorology, economics, and more. Here, we provide an explanation of deterministic dynamics and how it can be approximated for social phenomena.

Deterministic dynamics involves a mathematical approach to describe the behavior of systems obeying physical laws over time. This theory assumes that each element within the system, such as particles, variables, or agents, follows specific differential equations. These differential equations are dependent on initial conditions and predict the system's evolution over time.

Key features of deterministic dynamics include: - Reversibility: Systems governed by deterministic dynamics are reversible in time, allowing for the calculation of evolution from the past to the future. - Chaos: Some deterministic dynamics systems exhibit chaotic behavior, making predictions challenging even with small changes in initial conditions.

## 3.20 Approximating Social Phenomena Using Deterministic Dynamics

When approximating social phenomena using deterministic dynamics, several approaches can be considered:

1. Agent-Based Models: Social phenomena can be modeled by simulating the behavior of individual agents (individuals, organizations, elements, etc.) that interact based on specific deterministic rules. This approach captures the complex dynamics of society.

2. Differential Equation Models: Social phenomena can be modeled as a set of differential equations, describing how individual elements or parameters change over time. This approach is utilized in fields like demographics, infectious disease modeling, economics, and more.

3. Network Models: The relationships within society can be modeled as a network, applying deterministic rules to nodes and edges. This approach is often used in social network analysis.

4. Application of Chaos Theory: If social phenomena exhibit chaotic elements, chaos theory can be applied to investigate their dynamics and reveal the limits of predictability.

When applying deterministic dynamics to social phenomena, attention must be given to the nature of the system and the choice of models. Additionally, data collection and parameter estimation are critical steps. To capture the complexity of social phenomena, advanced mathematical tools and computer simulations are commonly employed.

## 3.21 Approximating Social Phenomena Using Torus

## 3.22 Introduction to Torus

A torus is an important geometric object in mathematics and physics. It has the shape of a donut, with an inner and outer circle that touch each other. Because a torus is periodic in both the circumferential and radial directions, it possesses special topological properties.

When approximating social phenomena as a torus, the following aspects can be considered:

1. Periodicity and Cycles: Due to its periodic shape, modeling social phenomena as a torus takes into account periodic trends and cyclic patterns. It is suitable for modeling phenomena with seasonal changes or cyclical behavior.

2. Inner and Outer Interactions: The inner and outer circles of a torus touch, allowing for the consideration of interactions between inner and outer elements. In the context of social phenomena, this can be applied to model interactions between different regions or elements.

3. Spatial Arrangement: Torus geometry is suitable for representing spatial arrangements. When modeling social phenomena with geographic considerations, the torus shape can be used to account for location information and distances.

4. Nonlinearity: Torus exhibits nonlinear properties. When social phenomena involve nonlinear relationships or interactions, using a torus can capture their complex dynamics.

When applying the torus to model social phenomena, it's crucial to leverage its characteristics and develop models that align with the nature of the phenomenon. Keep in mind that the specific situation and characteristics of the social phenomenon will determine how the torus is utilized.

## 3.23 Types of Attractors and Their Application to Social Phenomena

The following provides explanations of different types of attractors and how they can be applied to social phenomena.

## 3.24 Equilibrium Attractor

Description: An equilibrium attractor is an attractor where a system converges to one or more equilibrium points. Equilibrium points represent states in which the system remains stationary in the absence of external influences. Application to Social Phenomena: In social phenomena, equilibrium attractors can represent specific stable states or equilibrium conditions. For example, in economic models, when the market converges to a price where supply and demand are in balance, it can be considered an equilibrium attractor.

## 3.25 Periodic Attractor

Description: A periodic attractor is an attractor where a system oscillates in a specific periodic pattern. Periodic attractors repeat themselves at a specific period T. Application to Social Phenomena: In social phenomena, periodic attractors can be used to model periodic behaviors or events. For instance, modeling seasonal consumption patterns or economic cycles as periodic attractors is possible.

## 3.26 Quasi-Periodic Attractor

Description: A quasi-periodic attractor is similar to a periodic attractor but exhibits complex oscillations that are not strictly periodic. Application to Social Phenomena: In social phenomena, quasi-periodic attractors are useful for representing phenomena with periodicity that is not perfectly regular. For example, modeling daily fluctuations in stock prices or seasonal variations in weather as quasi-periodic attractors.

## 3.27 Strange Attractor

Description: A strange attractor is an attractor that exhibits complex, non-periodic oscillations in systems with nonlinear dynamics. It is a fundamental concept in chaos theory. Application to Social Phenomena: In social phenomena, strange attractors can be used to represent complex dynamics and unpredictable behaviors. For example, modeling price fluctuations in financial markets or congestion patterns in traffic flow as strange attractors.

The application of these attractors is employed in various fields such as social sciences, economics, meteorology, and traffic engineering, contributing to the understanding and prediction of complex phenomena. However, modeling social phenomena typically requires advanced mathematical techniques and data collection.

# 4. Study of mathematical models, Issues

# 5. Extensions of the Langford Equation in Opinion Dynamics

## 5.1 Introduction

When applying the Langford equation to opinion dynamics, various variations and combinations can be considered. This document discusses different extensions and adaptations of the Langford equation to incorporate additional elements.

## 5.2 Extended Langford Equation

The Langford equation can be extended by adding new terms to incorporate additional elements into the opinion dynamics. For example, introducing terms related to environmental factors or external stimuli can influence the opinion formation process.

## 5.3 Network Effects

Combining the Langford equation with network theory allows modeling interactions among agents. When agents are connected in a network and opinions propagate through the network, network effects can yield new insights.

## 5.4 Spatio-Temporal Effects

By adding spatial and temporal elements to the Langford equation, it becomes possible to model opinion formation processes that depend on geographical locations and the passage of time. This can account for regional differences in opinions and changes over time.

## 5.5 Agent Diversity

Introducing multiple types of agents with different attributes and opinions into the Langford equation allows for modeling interactions among diverse agents. This can capture the dynamics of a diverse population of agents.

## 5.6 Combination with Nonlinear Dynamics Models

Combining the Langford equation with compatible nonlinear models (e.g., Lotka-Volterra equation, Bernoulli equation, logistic equation, etc.) enables the modeling of complex dynamics. This approach enriches the modeling of interactions.

## 5.7 Incorporating Heuristics and Agent Behavior Rules

Incorporating agent heuristics and decision-making rules into the Langford equation allows modeling agents' rational behavior. Agent behavior patterns can influence opinion formation.

Based on the examples above, several mathematical models can be explored.

## 5.8 Consideration of Mathematical Models

In this section, we provide a proposed mathematical formula for incorporating spatiotemporal effects into the opinion dynamics model. The following formula represents spatiotemporal effects using the Laplacian (spatial gradient).

$$\frac{dx_i}{dt} = ax_i - bx_i^3 + c \sum_{j=1}^{N} W_{ij}(x_j - x_i)$$
$$+ dx_i(1 - x_i) + e \sum_{j=1}^{N} V_{ij} \tanh(x_j - x_i)$$
$$+ fMx_i + gx_i^2(1 - x_i) + \nabla^2 x_i$$

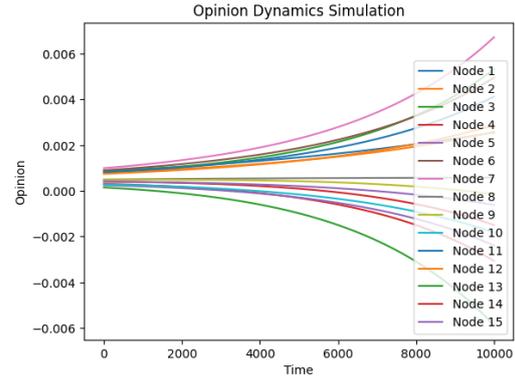

Fig. 7: Opinion Dynamics Simulation

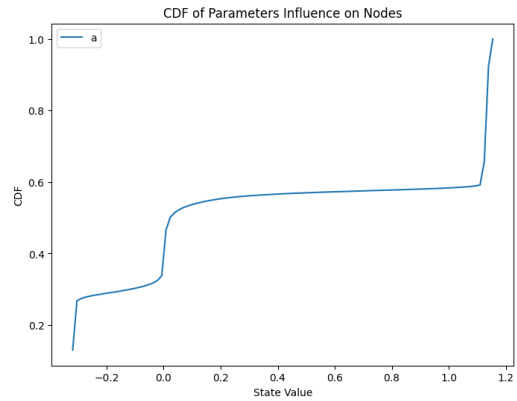

Fig. 8: Opinion Dynamics Simulation Parameta $a$

Here, $\nabla^2 x_i$ represents the Laplacian (spatial second derivative) of $x_i$. This allows the opinions of each individual $x_i$ to vary depending on their spatial location, taking into account spatiotemporal effects. The Laplacian influences the differences in opinions with neighboring individuals, potentially constraining the spatial propagation of opinions.

By using this formula for simulations, we can analyze in detail how opinions change over time and space, and how they spread. This approach allows us to incorporate spatiotemporal effects into the opinion dynamics model and explore the complex dynamics of the opinion formation process.

It's important to note that this proposed formula is just one example, and there are various ways to incorporate different spatiotemporal effects into the model.

## 5.9 Construction of an Opinion Dynamics Model Considering Spatiotemporal Effects and Regional Attributes

We present a proposed mathematical formula for constructing an opinion dynamics model that takes into account spatiotemporal effects and regional attributes. This model is applicable when opinion formation depends on both time and space and is influenced by regional attributes.

$$\frac{dx_i}{dt} = ax_i - bx_i^3 + c\sum_{j=1}^{N} W_{ij}(x_j - x_i)$$
$$+ dx_i(1 - x_i) + e\sum_{j=1}^{N} V_{ij}\tanh(x_j - x_i)$$
$$+ fM_i x_i + gx_i^2(1 - x_i)$$

In this equation, the newly introduced element is as follows:

- $M_i$: A parameter representing the regional attribute to which individual $i$ belongs. It is assumed that the influence on opinion formation varies based on regional attributes. This parameter may have different values for each region.

According to this model, individual opinion formation is influenced not only by interactions with other individuals but also by the regional attributes to which individuals belong. Therefore, by setting different regional attribute parameters for each region, regional differences and characteristics can be reflected in the opinion formation process.

Furthermore, this model can be simulated to investigate the opinion formation process under different regional attribute conditions. Regional attributes may lead to changes in the speed and patterns of opinion propagation.

### 5.10 Opinion Dynamics Model with Spatiotemporal Effects Considering Agent Distances

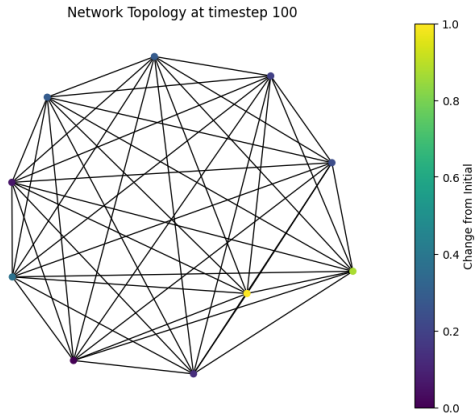

Fig. 9: Case:Network Topology

We present a proposed mathematical formula for an opinion dynamics model that incorporates spatiotemporal effects, taking into account the distances between agents. This model is applicable when the distances between agents influence the propagation of opinions.

$$\frac{dx_i}{dt} = ax_i - bx_i^3 + c\sum_{j=1}^{N} W_{ij}(x_j - x_i)$$
$$+ dx_i(1 - x_i) + e\sum_{j=1}^{N} V_{ij}\tanh(x_j - x_i)$$
$$+ fM_i x_i + gx_i^2(1 - x_i) + h\sum_{j=1}^{N} \frac{1}{r_{ij}}(x_j - x_i)$$

In this equation, the newly introduced elements are as follows:

- $r_{ij}$: A parameter representing the distance between agent $i$ and $j$. This is calculated based on the spatial positions of the agents and introduces an effect where interactions are strengthened between agents that are close in distance. - $h$: A weight parameter for distance. It adjusts the strength of the influence based on distance.

According to this model, the propagation speed and strength of opinions can vary depending on the distance between agents, and opinions of agents closer in distance may have a stronger influence on each other. By adjusting the model parameter $h$, you can control the strength of the distance effect.

Through simulations, it is possible to investigate the opinion formation process under different agent configurations and distance conditions. This may allow for a detailed analysis of the impact of agent distances on opinion formation.

### 5.11 Opinion Dynamics Model Incorporating Distance, External Influence, and Reliability ($D_{ij}$)

Understood. Here is the formula for an opinion dynamics model where distance, external influence, and reliability ($D_{ij}$) contribute interactively:

$$\frac{dx_i}{dt} = ax_i - bx_i^3 + c\sum_{j=1}^{N} W_{ij}(x_j - x_i) + dx_i(1 - x_i)$$
$$+ e\sum_{j=1}^{N} V_{ij}\tanh(x_j - x_i) + fM_i x_i + gx_i^2(1 - x_i)$$
$$+ h\sum_{j=1}^{N} \left(\frac{1}{r_{ij}}(x_j - x_i) + \alpha D_{ij}(x_j - x_i)\right)$$

In this equation:

- $r_{ij}$: A parameter representing the distance between agent $i$ and $j$, influencing the propagation speed of opinions based on distance. - $D_{ij}$: A parameter representing the reliability from agent $i$ to agent $j$, reflecting trust relationships between agents. - $\alpha$: A parameter controlling the strength of the impact of reliability ($D_{ij}$) on opinion propagation.

In this model, the effects of distance, external influence, and reliability interactively contribute to the opinion formation process. By adjusting various parameters ($h$, $\alpha$, etc.), you can control the influence of these factors.

This serves as an example of constructing the dynamics of opinion formation under scenarios with varying distance, external influence, and reliability through simulations.

## 5.12 Consideration of Topological Effects Using Opinion Dynamics Models and Network Theory in Network Formation

When forming a network considering topological effects using opinion dynamics models and network theory, the following steps are involved:

1. Network Construction: Build a network that represents the interactions between agents ($x_i$) in the model. The nodes of the network represent agents, and the edges represent relationships between agents. The weights of the edges are set based on interaction parameters $W_{ij}$ and $V_{ij}$.

2. Numerical Simulation: Conduct numerical simulations using the network. Calculate the propagation of opinions and dynamics between nodes (agents) over time. Update the opinion values ($x_i$) of each agent while evolving the equations over time.

3. Network Visualization: Create a network diagram from the simulation results. The network diagram consists of nodes and edges, and the positions of nodes are arranged based on the opinion values of agents. Edge thickness and colors can be used to represent the strength and types of interactions between agents.

4. Analysis of Topological Effects: Analyze topological effects in the opinion dynamics model using the network diagram. Evaluate network characteristics using network metrics such as node placement, edge properties, clustering, centrality, and more.

In this way, by combining numerical simulations with network theory, you can visualize and analyze the topological effects in opinion dynamics models. You can also explore the possibilities of understanding different patterns of opinion propagation under different parameter settings and network structures using the obtained results.

## 5.13 Integration of Polarization Elements from the Kertz Model into the Above Opinion Dynamics Model

To integrate the polarity elements of the Kertz model into the above opinion dynamics model, it is necessary to add the influence of polarity obtained from the Kertz model to the model. The following is an overview of the procedure:

1. Consideration of the Polarity Elements of the Kertz Model: The Kertz model includes parameters that represent the polarity of opinions. To incorporate these polarity elements into the above model, introduce new variables (e.g., $p_i$) that represent the polarity of each agent. $p_i$ represents the polarity of each agent and typically takes values in the range of -1 to 1.

2. Adding Influence from the Kertz Model: To integrate the polarity elements of the Kertz model into the model, add the influence of polarity to the equation for the change in opinion of each agent. Specifically, it can be modified as follows:

$$\frac{dx_i}{dt} = ax_i - bx_i^3 + c \sum_{j=1}^{N} W_{ij}(x_j - x_i) + dx_i(1 - x_i)$$

$$+ e \sum_{j=1}^{N} V_{ij} \tanh(x_j - x_i) + fMx_i$$

$$+ gx_i^2(1 - x_i) + hp_i$$

Here, $h$ is a parameter that adjusts the influence of polarity. If $h > 0$, polarity has a positive impact on opinion convergence, and if $h < 0$, polarity has a negative impact on opinion convergence.

3. Adjusting the Parameter $h$ for Influence from the Kertz Model: Adjust the value of the $h$ parameter to control the strength of the influence of polarity. This allows you to adjust the impact of polarity on opinion dynamics.

In this way, you can integrate the polarity elements from the Kertz model into the opinion dynamics model. The influence of polarity can be investigated, affecting agent opinion formation and the convergence of polarity.

## 5.14 Incorporating Weight Parameter $h$ for Distance

1. Introduction of Weight Parameter $h$ for Distance: To control the influence of distance between each agent, introduce a new parameter called $h$. This is a weight parameter for distance.

2. Addition of Distance Influence Term: Add a term for the influence of distance to the equation for the change in opinion of each agent in the model. Specifically, it can be modified as follows:

$$\frac{dx_i}{dt} = ax_i - bx_i^3 + c \sum_{j=1}^{N} W_{ij}(x_j - x_i) + dx_i(1 - x_i)$$

$$+ e \sum_{j=1}^{N} V_{ij} \tanh(x_j - x_i) + fMx_i + gx_i^2(1 - x_i) + h \sum_{j=1}^{N} D_{ij}(x_j - x_i)$$

Here, $D_{ij}$ represents a weight related to the distance between agent $i$ and agent $j$. When the distance is short, the value of $D_{ij}$ becomes large, and when it is far, it becomes small. In this way, the influence of distance can be adjusted through the weight parameter *h* for distance.

This results in an opinion dynamics model that takes into account the influence of distance. To investigate the impact of distance on opinion convergence, you can adjust the weights related to distance and the value of parameter *h*.

### 5.15 Incorporating Weight Parameter *h* for Distance in Kuramoto Model

To incorporate a weight parameter *h* for distance and the strength of distance-related influence into the Kuramoto model, you can make the following changes to the opinion dynamics model.

First, to account for the weight parameter *h* for distance, you add interaction terms weighted based on the distance between agents. Let $d_{ij}$ represent the distance between agent *i* and agent *j*, and $h_{ij}$ be the weight parameter for distance. Next, you add a term to control the strength of distance-related influence.

As a result, the update equation for the opinion dynamics model is modified as follows:

$$\frac{dx_i}{dt} = ax_i\left(1 - \frac{x_i}{K}\right) - \frac{\sum_{j=1}^{N} h_{ij} \cdot W_{ij} \cdot x_j}{1 + \sum_{j=1}^{N} h_{ij} \cdot x_j}$$

Here, $h_{ij}$ is the weight parameter calculated based on the distance $d_{ij}$ between agent *i* and agent *j*. By adjusting this parameter, you can control the strength of interaction based on distance.

In this way, you obtain an opinion dynamics model that considers the weight parameter *h* for distance and the strength of distance-related influence. The specific method for calculating the parameter $h_{ij}$ and adjusting the strength of influence can be customized to match the specific problem setting and data.

### 5.16 Incorporating Weight Parameter *h* and Distance-Related Influence in the Kuramoto Model within the Opinion Dynamics Model

To incorporate a weight parameter *h* for distance and the strength of distance-related influence into the Kuramoto model, you can make the following changes to the opinion dynamics model.

First, to account for the weight parameter *h* for distance, you add interaction terms weighted based on the distance between agents. Let $d_{ij}$ represent the distance between agent *i* and agent *j*, and $h_{ij}$ be the weight parameter for distance. Next, you add a term to control the strength of distance-related influence.

As a result, the update equation for the opinion dynamics model is modified as follows:

$$\frac{dx_i}{dt} = ax_i\left(1 - \frac{x_i}{K}\right) - \frac{\sum_{j=1}^{N} h_{ij} \cdot W_{ij} \cdot x_j}{1 + \sum_{j=1}^{N} h_{ij} \cdot x_j}$$

Here, $h_{ij}$ is the weight parameter calculated based on the distance $d_{ij}$ between agent *i* and agent *j*. By adjusting this parameter, you can control the strength of interaction based on distance.

In this way, you obtain an opinion dynamics model that considers the weight parameter *h* for distance and the strength of distance-related influence. The specific method for calculating the parameter $h_{ij}$ and adjusting the strength of influence can be customized to match the specific problem setting and data. In this paper, we base our discussion on this model.

## 6. Discussion

### 6.1 Proposal for Constructing an Opinion Dynamics Model

In the construction of an opinion dynamics model that takes into account magnetic field conditions, we propose an approach that combines the Langford equation, diffusion model, logistic map, and Katz model, and further incorporates elements of nonlinear dynamics.

### 6.2 Model Formulation

The following equations represent the proposed opinion dynamics model:

$$\frac{dx_i}{dt} = ax_i - bx_i^3 + c\sum_{j=1}^{N} W_{ij}(x_j - x_i)$$
$$+ dx_i(1 - x_i) + e\sum_{j=1}^{N} V_{ij}\tanh(x_j - x_i)$$
$$+ fMx_i + gx_i^2(1 - x_i)$$

Here,

$x_i$ represents the variable that expresses the opinion of individual *i*.

$a, b, c, d, e, f, g$ are model parameters.

$W_{ij}$ and $V_{ij}$ represent the weights of social interactions.

$M$ is a parameter that represents the strength of the magnetic field.

**Functions of Parameters**

$a$: Controls the basic growth rate of $x_i$.

$b$: Controls the nonlinear growth of $x_i$.

$c$: Adjusts the strength of social diffusion.

$d$: Controls the logistic growth rate.

*e*: Controls the strength of opinion polarization.

*f*: Represents the strength of the magnetic field's influence.

*g*: Controls the strength of the butterfly effect.

$W_{ij}$: Represents the strength of social interaction between agent $i$ and agent $j$.

$V_{ij}$: Represents the influence of the polarity of opinions between agent $i$ and agent $j$.

$M$: Adjusts the strength of the magnetic field.

The system is modeled using a network of nodes where each node's state is updated based on a differential equation. The parameters and initial conditions for the simulation are set as follows:

Number of nodes: $n = 10$

Fitting Parameters: $a = 0.002, \quad b = 0.0015, \quad c = 0.007,$
$$d = 0.003, \quad e = 0.004, \quad f = 0.007, \quad g = 0.09$$

Initial values for each node are set randomly within a small range:

$$X_0 = \text{Random value in } [0, 0.001)$$

Weight matrices $W$ and $V$ are initialized randomly:

$$W = \text{Random matrix with values in } [0, 0.25)$$
$$V = \text{Random matrix with values in } [0, 0.2)$$

Time settings for the simulation:

$$\text{Timesteps: } T = 100$$
$$\text{Time step size: } \Delta t = 0.01$$

The model function that describes the dynamics of each node is given by:

$$\frac{dx}{dt} = aX - bX^3 + c\sum W(X_i - X_j) + dX(1-X) + e\sum V \tanh(X_i - X_j) + fX + gX^2(1-X)$$

A function to calculate the influence of each parameter on the nodes is defined as follows. It identifies the parameter with the maximum influence at each timestep:

$$\text{Influence} = \operatorname{argmax}\Big(|aX|, |bX^3|, |c\sum W(X_i - X_j)|,$$

$$|dX(1-X)|, |e\sum V \tanh(X_i - X_j)|, |fX|, |gX^2(1-X)|\Big)$$

The following equations represent the proposed opinion dynamics model:

$$\frac{dx_i}{dt} = ax_i - bx_i^3 + c\sum_{j=1}^{N} W_{ij}(x_j - x_i)$$
$$+ dx_i(1 - x_i) + e\sum_{j=1}^{N} V_{ij} \tanh(x_j - x_i)$$
$$+ fMx_i + gx_i^2(1 - x_i)$$

Here,

$x_i$ represents the variable that expresses the opinion of individual $i$.

$a, b, c, d, e, f, g$ are model parameters.

$W_{ij}$ and $V_{ij}$ represent the weights of social interactions.

$M$ is a parameter that represents the strength of the magnetic field.

### Network Functions

*a*: Controls the basic growth rate of $x_i$.

*b*: Controls the nonlinear growth of $x_i$.

*c*: Adjusts the strength of social diffusion.

*d*: Controls the logistic growth rate.

*e*: Controls the strength of opinion polarization.

*f*: Represents the strength of the magnetic field's influence.

*g*: Controls the strength of the butterfly effect.

$W_{ij}$: Represents the strength of social interaction between agent $i$ and agent $j$.

$V_{ij}$: Represents the influence of the polarity of opinions between agent $i$ and agent $j$.

$M$: Adjusts the strength of the magnetic field.

visualizations of a network topology at different timesteps, with nodes colored according to the change from their initial state. The colors, represented on a scale from 0 to 1, indicate the magnitude of change, with the accompanying labels ('a', 'b', 'c', 'd', 'e', 'f', 'g') suggesting the influence of different parameters on each node.

Regarding the mathematical model and parameters, we can infer their roles as follows:

- *a*: Controls the basic growth rate of $x_i$. - *b*: Governs the nonlinear saturation effect on $x_i$. - *c*: Adjusts the strength of social diffusion. - *d*: Controls the logistic growth rate. - *e*: Controls the degree of polarization of opinions. - *f*: Represents the strength of an external field effect. - *g*: Adjusts the strength of the butterfly effect. - $W_{ij}$: Represents the strength of social interaction between agents $i$ and $j$. - $V_{ij}$:

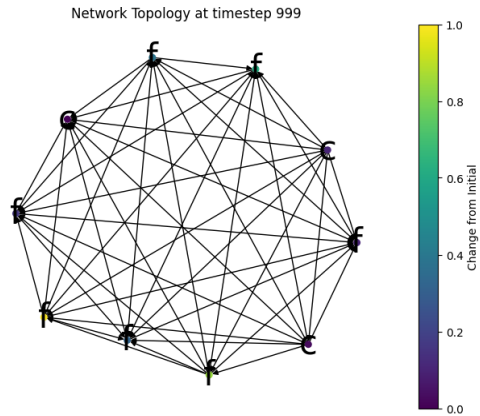

Fig. 10: Network Topology

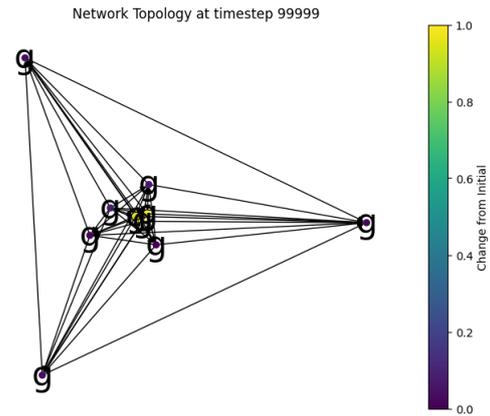

Fig. 12: Network Topology

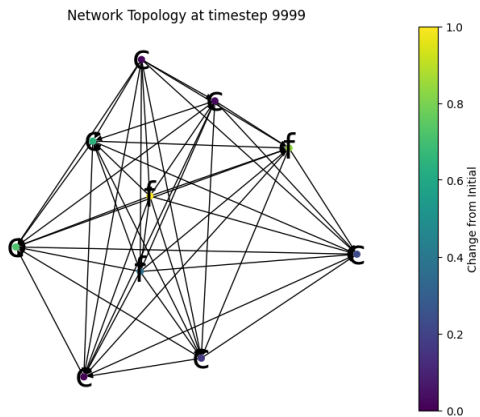

Fig. 11: Network Topology

Represents the influence of the polarity of opinions between agents $i$ and $j$. - $M$: Adjusts the strength of the magnetic field.

In the context of social phenomena, media influence, and consensus formation, these parameters could have the following interpretations:

### 1. Social Phenomena

Parameters like $c$ and $W_{ij}$ might model the degree to which individuals are influenced by their social network. This could reflect phenomena such as peer pressure, the spread of information, or the propagation of social norms.

### 2. Media Influence

The parameter $e$ could represent media influence if we consider the media's role in shaping public opinion and potentially polarizing views. $V_{ij}$ may also play a part here, indicating how different opinions are polarized by interactions.

### 3. Consensus Formation

The logistic growth rate $d$ and the saturation effect $b$ might be crucial in consensus formation, where opinions (or states) cannot grow indefinitely and tend to stabilize. The balance between conforming to the majority (consensus) and maintaining individual differences could be modeled by these parameters.

### Model and parameter $g$

The parameter $g$ in the model equation represents the non-linear feedback within the system and can have different interpretations depending on the context:

(1) Social Phenomena: $g$ models the disproportionate effects of individual actions or decisions, leading to sudden shifts in social dynamics.
(2) Media Influence: $g$ could represent the amplification of social tendencies by the media, potentially leading to polarized echo chambers.
(3) Consensus Formation: $g$ reflects the critical mass necessary for consensus formation, indicating a self-reinforcing spread of consensus beyond a certain agreement level.

The network topology at a significant timestep is visualized below, showing the influence of various parameters, including $g$, on the network's nodes.

## 6.3 CDF of Parameters Influence on Nodes

The result shows the Cumulative Distribution Function (CDF) of the state values for each node at the final timestep of the simulation. This type of graph helps to understand the distribution of states across the nodes at the end of the simulation period.

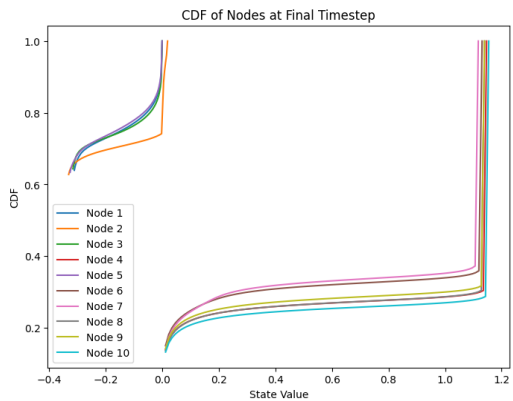

Fig. 13: CDF of Parameters Influence on Nodes

Given the parameters you've outlined and their influence on the model, let's discuss the considerations for social phenomena, media influence, and consensus formation:

(1) **Consideration of Social Phenomena:**

The distribution of state values at the final timestep can give insight into how social dynamics stabilize over time.

The parameter $g$, by controlling the strength of the butterfly effect, could indicate that small changes in individual behavior can lead to large differences in the final distribution of states, which is particularly relevant for understanding phenomena such as viral trends or the emergence of new norms within a society.

(2) **Consideration of Media Influence:**

Media plays a crucial role in influencing the state of nodes, which in this context could represent individual beliefs or opinions.

The CDFs might reflect the effectiveness of media in homogenizing opinions (if the CDFs are closer together) or in creating divergence (if the CDFs are spread out). The parameter $e$, controlling opinion polarization, could be a key factor in determining the shape of these CDFs.

(3) **Consideration for Consensus Formation:**

Consensus in a social network could be observed if the CDFs of all nodes converge to a narrow band, indicating that most nodes have similar state values.

The non-linear terms controlled by $b$ and $g$ could model the resistance or acceleration towards a consensus state. If the CDF shows a quick rise to 1, it could indicate rapid consensus formation.

These points could be further explored through simulations varying these parameters to see their effect on the final distribution of node states.

### 6.4 CDF of Parameters Influence on Nodes Parameter $D$ Representing Distance

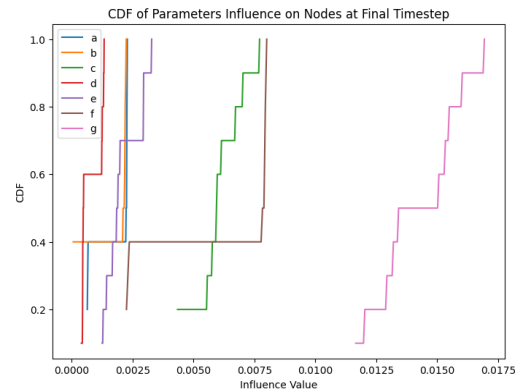

Fig. 14: CDF of Parameters Influence on Nodes at Final Timestep

The results have provided shows the Cumulative Distribution Function (CDF) of the influence values of various parameters on nodes at the final timestep of the simulation. This graph is particularly useful for analyzing the relative impact of each parameter on the nodes' state values throughout the simulation.

Let's discuss the difference of parameter $d$ in relation to other parameters from the perspectives of social phenomena, media influence, and consensus formation, building upon the conclusions from the CDF of nodes:

(1) **Consideration of Social Phenomena:**

Parameter $d$ might be related to the intrinsic growth rate of a node's state, possibly representing an individual's inclination to change independently of others. If $d$ shows a significant influence in the CDF, it suggests that personal growth or decay plays a critical role in the overall dynamics.

Compared to other parameters that may represent external influences or interactions, $d$'s effect could signify internal decision-making processes or intrinsic motivation within the social context.

(2) **Consideration of Media Influence:**

If media influence is modeled by parameters like $e$ or $f$, which might signify external forces or fields acting on nodes, $d$ stands out by possibly representing the natural receptivity or resistance of individuals to such influences.

The CDF of *d* could indicate the degree to which individual nodes are affected by media compared to their internal dynamics. A lower influence value for *d* would imply that external media influences are more decisive in shaping node states than intrinsic factors.

(3) **Consideration for Consensus Formation:**

In a model where consensus is reflected by a convergence of node states, the role of *d* could be seen in how quickly nodes reach a stable state, either by growing towards the consensus or resisting it.

A distinct pattern for *d* in the CDF compared to other parameters, particularly those modeling interactions like *c* and *g*, would suggest that the internal growth rate is a determining factor in whether and how nodes reach consensus.

These considerations of parameter *d* relative to others provide a nuanced view of how various factors contribute to the state dynamics within a networked system. It allows for the differentiation between the effects of internal versus external influences and the understanding of their respective roles in social behavior, media impact, and consensus-building processes.

## 6.5 Opinion Dynamics Simulation:Toroidal Polarization

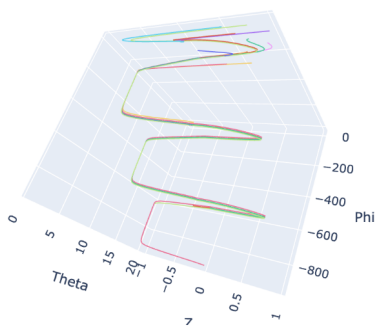

Fig. 15: Opinion Dynamics Simulation:Toroidal Polarization:2

**1. Social Phenomena Consideration:**

The trajectory plots might illustrate the dynamics of social interactions over time, with each curve representing a different social parameter or agent.

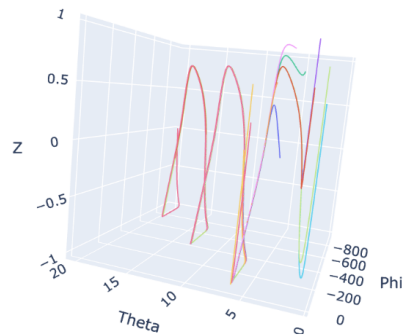

Fig. 16: Opinion Dynamics Simulation:Toroidal Polarization:2

Parameter differentials could indicate the varying influence of individual behavior (Theta) and social connectivity (Phi) on the overall system. For example, a steep curve in the plot could represent a sudden change in social behavior due to a critical event or threshold.

**2. Media Influence Consideration:**

These plots could represent the influence of media over time or across different segments of society. The Z-axis might denote the intensity of media influence, with the other axes representing different media strategies or content types.

The oscillations or patterns could signify how different media approaches resonate with the public, where closer loops may indicate a strong, repeated impact on public opinion or behavior.

**3. Consensus Formation Consideration:**

In terms of consensus formation, the plots could visualize the progression towards a common state or belief across a population. The convergence of lines might symbolize a move towards consensus, while divergence could represent disagreement or the persistence of diverse opinions.

The interaction of the Theta and Phi parameters with the Z-axis might show the role of internal (personal beliefs) and external (social pressure) factors in forming a consensus within a community or network.

## 6.6 Node of Toroidal Polarization

The provided Fig.16 displays multiple trajectories, each corresponding to a node in a networked system. The axes labeled

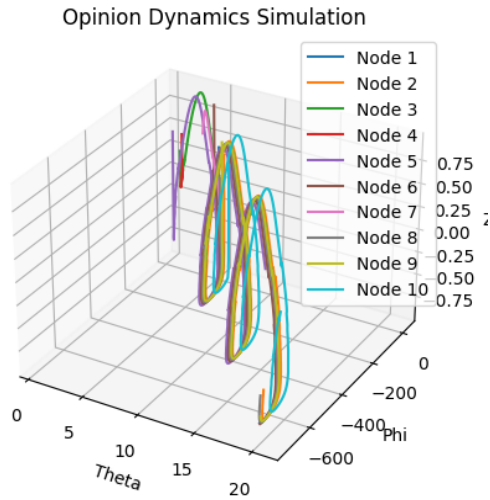

Fig. 17: Opinion Dynamics Simulation:Toroidal Polarization:3

Theta, Phi, and Z represent variables or parameters within the opinion dynamics model.

1. **Social Phenomena Consideration:**

    The trajectories suggest dynamic changes in opinions over time within a social network. The variation in the paths of different nodes indicates diversity in response to social stimuli.

    Theta could represent personal conviction or stubbornness, affecting how strongly an individual adheres to their initial opinion despite social pressure (Phi). The plot shows that individuals (nodes) may experience swings in their opinions before settling, which could represent real-life scenarios of changing social stances.

2. **Media Influence Consideration:**

    The nodes' trajectories could also reflect the varying impact of media on public opinion. Nodes that exhibit similar patterns may be influenced similarly by media campaigns or news cycles.

    The Z-axis might quantify the degree of media influence, with each node's fluctuation representing the changing impact on that individual's opinion. Large swings in the Z-axis for particular nodes could suggest susceptibility to media influence, whereas minimal change might indicate resistance.

3. **Consensus Formation Consideration:**

    The convergence or divergence of node trajectories can imply the formation of consensus or persistent disagreement within the network. If nodes converge to a similar endpoint, it could suggest that a consensus has been reached.

    The plot may show that while some nodes quickly align with the consensus (showing less deviation in their trajectories), others may take a more circuitous route, reflecting individual differences in the process of reaching a common stance.

    These interpretations, inferred from the Opinion Dynamics Simulation plot, underscore the complexity of modeling opinion dynamics. The visual representation helps in understanding the multifaceted influences on individual and collective opinion formation, the role of external factors such as media, and the pathways through which a group may reach a consensus.

### 6.7 CDF of Parameters $Theta$, $Phi$, $Z$, $W$

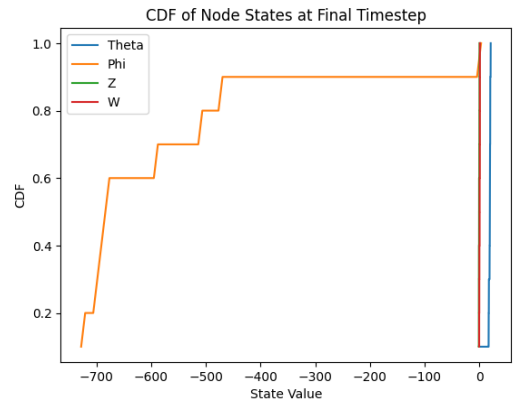

Fig. 18: CDF of Parameters Influence on Nodes at Final Timestep

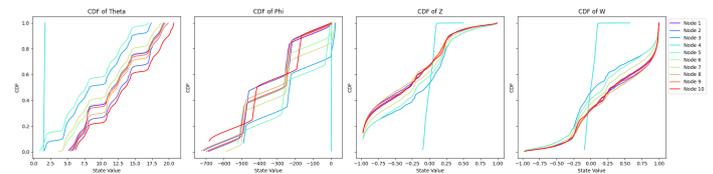

Fig. 19: CDF of Parameters $Theta$, $Phi$, $Z$, $W$

Results show the Cumulative Distribution Function (CDF) for four different parameters—Theta, Phi, Z, and W—across various nodes in an opinion dynamics simulation at the final timestep.

1. **Social Phenomena Consideration:**

    The CDF of Theta could represent individual tendencies within the network, with the spread of the CDF indicating a diversity in the population's opinion resilience or adaptability.

The CDF of Phi, showing multiple step-like progressions, may indicate clusters of nodes that have similar levels of influence or are influenced similarly by their neighbors, reflecting social segments or echo chambers within the network.

## 2. Media Influence Consideration:

The Z parameter's CDF might measure the intensity of external factors such as media or propaganda, with the curve's shape reflecting how opinions are swayed or polarized by such influences. A steep rise in the CDF could point to a strong, possibly uniform impact of media across nodes, while a more gradual slope could suggest varied media impact across different nodes.

The CDF of W could potentially relate to external factors beyond media, such as socioeconomic or cultural influences that affect opinion formation. The variation in the curves between nodes could indicate heterogeneity in how these factors play a role in shaping individual opinions.

## 3. Consensus Formation Consideration:

If consensus is represented by nodes reaching similar state values, the convergence of CDFs at higher state values could indicate nodes moving towards a common opinion or decision.

Conversely, the divergence in CDFs, especially if seen in the Phi or W parameters, might imply that despite commonalities in some aspects of opinion formation (like susceptibility to media), consensus is not achieved due to other diverging influences or stubbornness in personal belief (Theta).

These interpretations, deduced from the CDF plots, highlight the multi-dimensional and complex nature of opinion dynamics within a network. They also emphasize the importance of considering various factors, both internal (Theta) and external (Phi, Z, W), when analyzing social consensus formation, media influence, and overall social dynamics.

# 7. Conclusion

## 7.1 Issues with the Proposed Opinion Dynamics Model

1. **Abundance of Parameters**: - The proposed model contains numerous parameters ($a, b, c, d, e, f, g, W_{ij}, V_{ij}, M$). Selecting and tuning these parameters is challenging and often requires extensive experimentation to fit real-world data. Excessive parameters can increase model complexity and the risk of overfitting.

2. **Clarification of Parameter Meanings**: - It's crucial to provide clear explanations of the meanings and impacts of each parameter. There is a need for explanations regarding what each parameter represents and how it influences the model.

3. **Complexity of Nonlinear Dynamics**: - The model includes nonlinear dynamics, which can lead to complex behaviors. Understanding and predicting nonlinear dynamics can be challenging and may pose interpretability challenges for the model.

4. **Difficulty in Parameter Tuning**: - Parameter tuning is highly challenging, requiring extensive experiments to find suitable values. Moreover, different combinations of parameters may lead to different outcomes, making it difficult to understand their effects.

5. **Comparison with Real Data**: - Evaluating the model's validity requires comparison with real data. However, acquiring and aligning real data with the model can be challenging, necessitating reliable data sources.

6. **Model Interpretability**: - Due to the model's complexity, interpreting its results can be challenging. Sufficient interpretability should be incorporated into the model's outcomes and explanations.

7. **Computational Resources**: - The model demands significant computational resources, particularly when conducting large-scale simulations. Attention to computational time and memory constraints is essential.

To overcome these challenges and make the model practical and useful, careful parameter tuning, data collection, and comparison with real data, as well as improving model interpretability, are necessary.

# Aknowlegement

The author is grateful for discussion with Prof. Serge Galam and Prof.Akira Ishii. This research is supported by Grant-in-Aid for Scientific Research Project FY 2019-2021, Research Project/Area No. 19K04881, "Construction of a new theory of opinion dynamics that can describe the real picture of society by introducing trust and distrust".